\documentclass[aps,pra,superscriptaddress,twocolumn,balancelastpage]{revtex4-1}
\usepackage{amsmath}
\usepackage{amssymb}
\usepackage{graphicx}
\usepackage{bm}
\usepackage{color}
\usepackage{natbib}
\setcitestyle{numbers}
\setcitestyle{square}
\usepackage[colorlinks=true,
urlcolor=blue,
linkcolor=blue,
citecolor=blue]{hyperref}

\usepackage{url}
\usepackage{hyperref}

\newcommand{\kt}[1]{\left \lvert #1 \right \rangle}

\DeclareMathOperator{\tr}{tr}

\newcommand{\rev}[1]{\textcolor{black}{#1}}

\graphicspath{{../figures/}}
\newcommand{\unity}{1\!\!1}

\begin{document}

\title{Sunburst quantum Ising battery}
\author{Akash Mitra}
\author{Shashi C. L. Srivastava}
\affiliation{Variable Energy Cyclotron Centre, 1/AF Bidhannagar, Kolkata 
	700064, India}
\affiliation{Homi Bhabha National Institute, Training School Complex, 
	Anushaktinagar, Mumbai - 400094, India}
\begin{abstract}
    We study the energy transfer process in the recently proposed sunburst 
    quantum Ising model, which consists of two interacting integrable 
    systems: a transverse Ising chain with a very small transverse field 
    and a finite number of external isolated qubits. We show that in this 
    model of the quantum battery, coupling between the battery and charger 
    can be used to optimize the ergotropy, which is the maximum amount of 
    energy that can be extracted from the battery. At the same time, 
    maximum charging power increases with the coupling strength, allowing 
    for the simultaneous optimization of both ergotropy and charging power 
    in the strong coupling limit. Furthermore, we show that both ergotropy 
    and charging power are independent of the initial state of the 
    charger. 
\end{abstract}
\maketitle
\section{Introduction}

With the recent advances in quantum technology, there is a new stimulus 
for 
theoretical questions about thermodynamic principles, heat exchange, 
work extraction, etc, in the quantum regime. Small quantum systems that 
can store the energy at the point of source and deliver the energy to 
desired quantum systems are called quantum batteries (QBs), a term 
coined by Alicki and Fannes while studying the maximal amount of work 
that can be extracted from a quantum system\cite{Alicki_2013}. Since 
then, several studies have explored different aspects of quantum 
batteries such as charging schemes, work extraction, charging speed, the 
role of correlations, many-body interaction etc. \cite{Binder_2015, 
Campaioli_2017,Hovhannisyan_2013,Andolina_2019,Pirmoradin_2019,Ferraro_2018,Crescente_2020,Zhang_2023,Marcello_2018,Le_2018,Ghosh_2020,Zhao_2021,Rossini_2019,Ghosh_2021,Zhang_2019,Liu_2021,Peng_2021,Hu_2022}.
 While entangling control operations has been shown 
favorable for maximum work extraction \cite{Alicki_2013}, entanglement 
between charger and battery or different parts of battery has been argued 
to reduce the ergotropy \cite{Andolina_2019}. However, this 
can be handled to a certain extent by optimization over the initial state 
of the charger \cite{Andolina_2019}. Similar initial state dependence of 
the 
charger for maximal work extraction in central spin quantum battery has 
been shown in \cite{Liu_2021}. In a recent experimental study, coherence 
has been identified as the key resource for the charging process 
\cite{HuaWan2023}. The quantum advantage in terms of charging 
power has been shown in the collective charging scheme in contrast to 
the parallel charging scheme \cite{Binder_2015, 
Campaioli_2017,Hovhannisyan_2013}. The 
collective charging scheme has been realized in one set of models by 
charging two-level systems (QBs) inside a single optical cavity 
\cite{Andolina_2019,Pirmoradin_2019,Ferraro_2018,Crescente_2020,Zhang_2023,Marcello_2018},
 while other set of models have realized the collective nature by 
allowing interaction between different batteries 
\cite{Le_2018,Ghosh_2020,Zhao_2021,Rossini_2019,Ghosh_2021,Zhang_2019}. 
Collective charging schemes enhance the charging power in various 
spin-chain battery models. Defects and impurities in disordered spin 
chains enhance disorder-averaged power \cite{Ghosh_2020} but ergotropy 
decreases when these disordered spin chains are in many-body localized 
regime \cite{ArjMohSantos2023}.

The role of multiple chargers and the number of batteries in the 
power stored per unit battery have been explored using multiple central 
spin chain batteries. In the two limits, first, when the number of 
chargers is more as compared to the number of batteries $N_B$, the power 
stored per unit battery increases proportional to $\sqrt{N_B}$ while in 
the second limit of the number of batteries being comparable with the 
number of chargers, the power stored per unit battery decreases as 
${N_B}^{-1/2}$ \cite{Peng_2021}. Further, it has been shown that maximum 
ergotropy for the central spin battery strongly depends on the initial 
state of chargers \cite{Liu_2021}. Recently, this model of quantum 
battery has been studied experimentally in NMR systems \cite{Joshi_2022} 
and has shown the $\sqrt{N}$ advantage when $N$ number of chargers are 
used in the collective scheme as first shown theoretically in 
\cite{Peng_2021}.

Several questions naturally arise, like whether it is possible to 
devise a scheme where ergotropy (first defined in \cite{Balian_2004}) 
and charging power are almost independent of the initial state of the 
charger. For a fixed initial state, it has been shown that charging 
power increases with increasing coupling between charger and battery at 
the expense of lower stored energy \cite{Ferraro_2018, Andolina_2018}. 
Is it possible to optimize charging power and ergotropy simultaneously 
by tuning the coupling strength? To explore these questions, we study 
the sunburst quantum Ising model as a quantum battery.

This paper is organized as follows: Section \ref{sec:exact_soln} 
introduces the sunburst quantum Ising battery and analytical solution of 
its time-evolved state for a single battery. We analyze the performance 
of 
the sunburst quantum Ising battery by calculating different figures of 
merit such as ergotropy, charging power, and charging time analytically 
for both single and two batteries, and numerically for more than two 
batteries in Section \ref{sec:battery_performance}. The initial state 
independence of both ergotropy and charging power is discussed in 
Section \ref{sec:in_st_independence}. Finally, we present the summary in 
Section \ref{sec:summary}.

\section{Sunburst quantum Ising battery}\label{sec:exact_soln}
A recently proposed model called the sunburst quantum Ising model has been 
studied from the perspective of quantum phase transitions 
\cite{Franchi_2022}, out of equilibrium dynamics in weak coupling limit 
\cite{Rossini_2022} and strong coupling limit \cite{Akashash_2023}. We 
re-imagine this model as an extended charger modeled by a transverse field 
Ising model with periodic boundary condition and various qubits attached 
to it as batteries. The Hamiltonian of this model is,
\begin{equation}
    H = H_c \otimes \mathbb{I}_{b} + \mathbb{I}_{c} \otimes H_b + V_{cb},
\end{equation}
where $H_c (H_b)$ is the Hamiltonian of the extended charger (batteries), 
$\mathbb{I}_{b(c)}$ is the identity operator in the space of batteries 
(charger), and $V_{cb}$ represents the interaction term between charger 
and batteries. The Hamiltonians for individual subsystems and the 
interaction Hamiltonian are defined as follows:
\begin{equation}
    \begin{aligned}
		H_c&=-\sum_{i=1}^L \left(J\sigma_i^x \sigma_{i+1}^x 
		+h\sigma_i^z\right)\\
		H_b &= -\frac{\delta}{2}\sum_{i=1}^n \Sigma_i^z; 
		\quad 
		{\color{black}	V_{cb} = -\kappa \sum_{i=1}^n \sigma_{1+(i-1)d}^x 
			\Sigma_{i}^x}.
	\end{aligned}
\end{equation}

Here, $L$ represents the total number of lattice points in the Ising 
ring 
acting as an extended charger ($J>>h$), and $\sigma_i$ denotes Pauli 
matrices on the $i$-th site, with periodic 
boundary conditions implying $\sigma_{L+1}^x = \sigma_1^x$. Without loss 
of generality, we assume $J > 0$ to ensure ferromagnetic interaction 
between spins, and $h$ is the strength of the transverse field. The spin 
chains are externally coupled to $n$ batteries (qubits), described by 
the 
Hamiltonian $H_b$, where $\Sigma_i$ denotes the Pauli matrix 
corresponding to the $i$-th battery. The energy gap between the two 
lowest eigenstates for the battery is denoted by $\delta$ and $\kappa$ 
represents the strength of interaction between the charger and 
batteries. 
Additionally, $d$ represents the distance between consecutive batteries. 

One of the most striking features of this model is that the total 
bipartite system, which, in the presence of interaction between the two 
subsystems, is no longer integrable, even though the individual 
subsystems are integrable \cite{Akashash_2023}. The non-integrability of 
the sunburst quantum Ising model contrasts with the central spin model, 
which is always in the integrable regime \cite{Villazon_2020}. 
However, the quantum chaotic nature of the system only becomes apparent 
when all the parameters of the Hamiltonian are comparable. \rev{As 
batteries are locally coupled to a charger, 
charging multiple batteries becomes easier than the central spin model, 
where each battery is connected globally.}
Before we present the numerical results of the energy transfer process 
for multiple batteries, we first analyze the charger-battery system 
within the limit of a single battery.

Initially, the charger and batteries are decoupled from each other and 
prepared in the state $\kt{\psi^I_G} = \frac{1}{\sqrt{2}}[ 
\kt{+++\dots+} 
+ \kt{---\dots-}] $ and $\kt{0}$ respectively with $\sigma_x \kt{\pm} = 
\pm \kt{\pm}, \Sigma_z \kt{0} = \kt{0}$. $\kt{\psi^I_G}$, a cat state 
\cite{Schrodinger1935}, or Greenberger-Horne-Zeilinger state 
\cite{Greenberger1989}, is the ground state of the transverse Ising 
chain to a very good approximation in the limit $J>>h$. Later on, we 
will see that this particular choice of the charger's initial state does 
not affect the energy transfer between the battery and the charger. The 
charging process starts at $t\geq 0$ when the charger and battery 
interact with the term $V_{cb}$. The time evolved state at arbitrary 
time $t$ is given by, 	
\begin{equation}\label{eq:coeff_psi_t_j0}
	\begin{aligned}
	\kt{\psi(t)} &= A(t)\kt{\psi^I_G}\otimes \kt{0}+B(t) \kt{\psi^I_N} 
	\otimes\kt{1}, \\
	\kt{\psi^I_N}&= \frac{1}{\sqrt{2}}[ \kt{+++\dots+} - \kt{---\dots-}]\\
	A(t) &= \cos \frac{\omega t}{2}+i\frac{ \delta}{\omega}\sin 
	\frac{\omega 
				t}{2}, B(t) = \frac{2i\kappa}{\omega}\sin \frac{\omega 
				t}{2},
	\end{aligned},
\end{equation} 
where $\omega = \sqrt{\delta^2 + 4\kappa^2}$ as obtained in 
\cite{Akashash_2023}. Notably, $\kt{\psi^I_N} 
\otimes \kt{1}$ also is an eigenstate of the Hamiltonian of decoupled 
charger and battery with eigenvalue $-J+\frac{\delta}{2}$.

Having the exact time evolved state at arbitrary time $t$ 
(Eq.\ref{eq:coeff_psi_t_j0}), in the next 
section, we study the characteristics of sunburst quantum Ising batteries 
in terms of charging time, entanglement, and ergotropy.

\section{ Entanglement and Ergotropy}\label{sec:battery_performance}
We use linear entropy to quantify entanglement between the charger and 
battery, which, more accurately is a measure of mixedness
\rev{that, in turn, increases with increasing entanglement between the 
subsystems. In other words, linear entropy will always be non-zero for a 
bi-partite entangled state. Being simple to calculate (with no 
requirement for diagonalization of the reduced matrix) and, at the same 
time, experimental accessibility \cite{islam2015, kaufman2016, 
linke2018measuring, Roos2019renyimeasure} makes it an attractive 
quantity to study.}
The linear entropy for the battery is defined as:
\begin{equation}
(S_L)_b = 1 - \text{tr}(\rho_b^2),
\end{equation}
where $\rho_b$ is the reduced density matrix of the battery obtained by 
tracing over the charger degrees of freedom of the density matrix of the 
complete system. While charging, charger and battery at time $t$ are in 
state $\kt{\psi(t)}$ given in Eq. \ref{eq:coeff_psi_t_j0}. The linear 
entropy for the battery is given by\cite{Akashash_2023}:
\begin{equation}
\begin{aligned}
\label{eq:Linear_entropy}
    S_L &= 1 - \left( \left[\cos^2\left(\frac{\omega t}{2}\right) + 
    \frac{\delta^2}{\omega^2} \sin^2\left(\frac{\omega 
    t}{2}\right)\right]^2 \right.\\
    &\quad + \left. \left[\frac{4 \kappa^2}{\omega^2} 
    \sin^2\left(\frac{\omega t}{2}\right)\right]^2 \right).
\end{aligned}
\end{equation}

With the introduction of coupling between the battery and charger, the 
energy of the battery is increased from its initial ground state energy 
$E(0)$. The energy stored in the battery at time $t$ is given by
\begin{equation}\label{eq:storedE_def}
    \Delta E(t)=E(t)-E(0),
\end{equation}
where $E(t)=\tr(\rho_b(t)H_b)$.
However, not all of the stored energy in the battery can be extracted in 
the form of thermodynamical work. The maximum amount of energy that can be 
extracted is known as ergotropy, and is defined as \cite{Balian_2004}
\begin{equation}\label{eq:ergo_def}
    \xi(t)=E(t)-E_p(t)
\end{equation}
where $E_p(t)=\tr(\rho^p_b(t) H_b)$ represents the energy of the passive 
state $\rho^p_b(t)$ in the space of battery. In the 
eigenbasis of the Hamiltonian $H_b$, 
the passive state is defined as \cite{Pusz1978,Lenard1978}
\begin{equation}
\rho^p_b(t) = \sum_{j=0}^{2^n-1}|c_j(t)|^2 |j\rangle\langle j|
\end{equation}
where $|c_j(t)|^2$ is occupation probability of $j$th eigenstate $\kt{j}$ 
of 
$H_b$, and $|c_j(t)|^2 \geq |c_{j+1}(t)|^2$. No energy can be extracted 
from the passive 
state since the lower energy states are always more populated than the 
higher energy states. 

For a single battery, the two possible states denoted as $|0\rangle$ and 
$|1\rangle$, are associated with probabilities $|A(t)|^2$ and $|B(t)|^2$. 
The ergotropy is non-zero only when $|B(t)|^2 \geq |A(t)|^2$, leading to 
the following constraint on the parameters:
\begin{align}\label{eq:kap_condition}
|B(t)|^2 \geq |A(t)|^2 \implies 2\kappa \geq \delta.
\end{align}
This constraint arises naturally, signifying that the battery charging is 
a consequence of the interaction. Therefore, the strength of the 
interaction must be sufficiently high to ensure that the excited state 
probability surpasses that of the ground state. Failure to meet this 
condition would result in the battery remaining predominantly in the 
ground state, precluding any work extraction.

The time (say $t^*$) at which $|A(t^*)|^2$ and $|B(t^*)|^2$ are equal is,
\begin{equation}\label{eq:charging_begin_time}
    t^*=\frac{2}{\omega}\cos^{-1}\left[\pm\sqrt{\frac{ 
    4\kappa^2-\delta^2}{8\kappa^2}}\right].
\end{equation}
  The ergotropy remains non-zero for time $t$ between $t_1^*$ and  
  $t_2^*$, as during this period, $|B(t)|^2 \geq |A(t)|^2$. 
  Here, $t_1^*$ 
  ($t_2^*$) corresponds to the time associated with the positive 
  (negative) sign inside the inverse cosine function.
\begin{figure}
    \centering
    \includegraphics{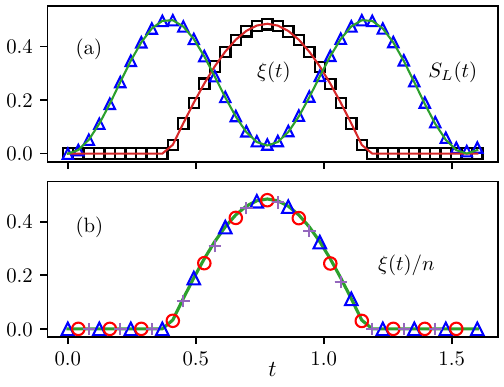}
    \caption{(a) The linear entropy and ergotropy are plotted for a 
    single battery with an extended charger of length $L=11$. The 
    numerical data points for linear entropy and ergotropy are denoted 
    by blue triangles and black squares, respectively. The solid green 
    line and the red line correspond to the analytical formulas derived 
    in Eq.\ref{eq:Linear_entropy} and Eq. \ref{eq:erg_exp}, 
    respectively. (b) The ergotropy per unit battery is plotted for 
    different numbers of batteries. The system sizes are selected as 
    $L=10, n=2$ (blue triangles), $L=9, n=3$ (red circles), and $L=8, 
    n=4$ (purple plus). The collapse of all the data on a single line 
    demonstrates the linear scaling of ergotropy with number of 
    batteries. In both figures, the parameters of the Hamiltonian are 
    chosen as $J=1$, $h=0.1$, $\delta=0.5$, and $\kappa=2$.}
    \label{fig:le_erg_vs_time_one_qubit}
\end{figure}
We obtain the ergotropy $\xi(t)$ using Eq. \ref{eq:ergo_def} as
\begin{equation}\label{eq:erg_exp}
    \xi(t)= 
    \begin{cases}
    \delta \left(|B(t)|^2-|A(t)|^2\right) & \text{ $t_1^* \leq t < 
    t_2^*$}\\
    0 & \text{otherwise}
    \end{cases}.
\end{equation}
Substituting the values of $|A(t)|^2$ and $|B(t)|^2$  from  
Eq. \ref{eq:coeff_psi_t_j0} in the above equation, we get the nonzero 
ergotropy as 
\begin{equation}\label{eq:erg_exp_final}
    \xi(t)=\delta\left[\left(\frac{8\kappa^2}{4\kappa^2+\delta^2}\right)\sin^2\frac{\omega
     t}{2}-1\right].
\end{equation}
The above analytical expression of ergotropy is compared with the 
numerical result obtained by exact diagonalization of sunburst quantum 
Ising battery in Fig. \ref{fig:le_erg_vs_time_one_qubit} and found in 
complete agreement. The stored energy in the battery at time $t$ as 
defined in Eq. \ref{eq:storedE_def} is given by
\begin{equation}\label{eq:stored_eng}
\Delta E(t) = 
\delta\left[\left(\frac{4\kappa^2}{4\kappa^2+\delta^2}\right)\sin^2\frac{\omega
 t}{2}\right].
\end{equation}
The amount of energy which is unavailable for extraction is given by the 
difference between stored energy and ergotropy and is expressed as
\begin{equation}
\begin{aligned}
    E_{\text{unavailable}}=\Delta 
    E(t)-\xi(t)=\delta\left(1-|B(t)|^2\right)\\=\delta\left[1-\left(\frac{4\kappa^2}{4\kappa^2+\delta^2}\right)\sin^2\frac{\omega
     t}{2}\right].
\end{aligned}
\end{equation}
Since $0 \leq |B(t)|^2 < 1$, the amount of unavailable energy is always 
positive consistent with the results obtained for other models of quantum 
batteries \cite{Andolina_2019,Kamin_2021,Andolina_2018}. For large 
interaction strength $\kappa$, $\max(|B(t)|^2) \to 1$, and therefore, the 
unavailable energy vanishes. The charging time ($T$) is 
minimum time at which stored energy becomes maximum during the charging 
process and, in this case, is given by,
\begin{equation}\label{eq:charging_time}
T = \frac{\pi}{\sqrt{\delta^2+4\kappa^2}}.
\end{equation}

At charging time, we see that along with stored energy, ergotropy attains 
its maximum, while the linear entropy is at its minimum non-zero value 
(see Fig. \ref{fig:le_erg_vs_time_one_qubit}). This agrees with 
earlier results for central spin battery \cite{Liu_2021}.

We now analyze the performance of the sunburst quantum Ising battery in 
the presence of multiple batteries and a single extended charger. We begin 
our discussion with two battery cells, considering the initial state as 
$\kt{\psi(0^-)} = \kt{\psi_G^I} \otimes \kt{00}$. In the case of two 
battery cells, we can analytically derive the exact expressions of the 
probabilities for the batteries to be in states $\kt{00}$ and $\kt{11}$. 
These two probabilities are sufficient to obtain the exact analytical 
expressions for both the stored energy and ergotropy since the Hamiltonian 
$H_b$ is diagonal with the elements $\text{diag}\left(-\delta, 0, 
0,\delta\right)$. We find that both the stored energy and ergotropy are 
exactly twice the stored energy and ergotropy given in  Eq. 
\ref{eq:stored_eng} and Eq. \ref{eq:erg_exp_final} respectively obtained 
for the case of a single battery. The details of the calculation are 
presented in the Appendix. \ref{ap:two_qubits}. 

We have numerically calculated ergotropy using the exact diagonalization 
method for multiple numbers of batteries. To compare these numerical 
results, we plotted ergotropy per battery and the analytical results 
obtained for single and double batteries in Fig. 
\ref{fig:le_erg_vs_time_one_qubit}. Encouraged by a complete collapse of 
all the numerical data on analytical result (Eq. 
\ref{eq:erg_exp_final}) for ergotropy, we conjecture that ergotropy 
increases linearly with the number of batteries. 
\begin{equation}
\xi(t) \propto n.
\end{equation}
However, the charging time remains the same as that of a single battery, 
implying that the charging time is independent of the number of batteries 
in agreement with the earlier work for the central spin battery 
\cite{Liu_2021}.

\begin{figure}[h]
    \centering
     \includegraphics{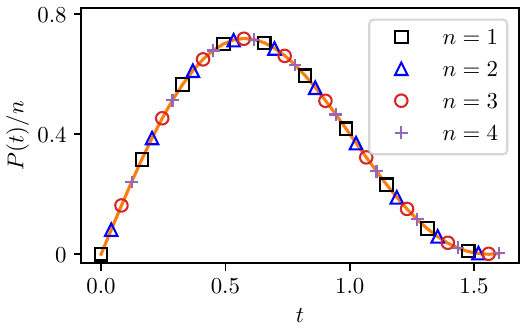}
    \caption{Charging power per unit battery as a function of time is 
    plotted for different numbers of batteries. A collapse of all the 
    data  
    reveals the linear scaling of charging power with the number of 
    batteries. 
    The system sizes are selected as $L=11,n=1$ (black squares), $L=10, 
    n=2$ (blue triangles), $L=9, n=3$ (red circles), and $L=8, n=4$ 
    (purple plus). The solid orange line corresponds to the analytical 
    formula derived in Eq.\ref{eq:power_one_qubit} for single battery 
    cell.}
    \label{fig:power_vs_time_multiple_qubits}
\end{figure}

The other figure of merit for quantum batteries is the charging power of 
the battery, which is defined as,
\begin{equation}\label{eq:power_def}
    P(t)=\frac{\Delta E (t)}{t},
\end{equation}
where $\Delta E(t)$ is stored energy defined in Eq. \ref{eq:storedE_def} 
and calculated for sunburst quantum Ising battery in Eq. 
\ref{eq:stored_eng}. The charging power then can be directly obtained as
\begin{equation}\label{eq:power_one_qubit}
    P(t)=\frac{\delta}{2}\left[ 
    \frac{8\kappa^2}{4\kappa^2+\delta^2}\frac{\sin^2\left(\frac{\omega 
    t}{2}\right)}{t}\right],
\end{equation}
which goes to 0 linearly in $t\to 0$ while for large time $P(t)$ 
approaches 0 as $t^{-1}$. Numerically, we calculate the charging power 
using the exact 
diagonalization method for different numbers of batteries. For a single 
battery, the numerical result matches well with the analytical result 
obtained in Eq. \ref{eq:power_one_qubit} (see Fig. 
\ref{fig:power_vs_time_multiple_qubits}).
For more than one battery cell, we observe a linear scaling of stored 
power with the number of battery cells (see Fig. 
\ref{fig:power_vs_time_multiple_qubits}), similar to the ergotropy. 
Specifically, we find that $P_{\text{max}} \propto n$ for a 
single extended charger. This is in contrast to the central spin model, 
where $P_{\text{max}} \propto \sqrt{n}$ in the limit where number of 
charger approaches one \cite{Peng_2021}. Consequently, the sunburst 
quantum 
Ising battery 
exhibits a charging rate that is $\sqrt{n}$ times faster compared to the 
central spin battery. A superior scaling of $n^{3/2}$ is observed for 
Dicke-type batteries where the extra $\sqrt{n}$ comes from the fact that 
charging time reduces as $n^{-1/2}$ 
\cite{Ferraro_2018,Zhang_2023,Crescente_2020}.

The power is maximized at time $t_{\text{max}} \approx 2.3312/\omega$ and 
the maximum 
power is 
\begin{equation}\label{eq:max_power}
    P_{\max}\left(t=t_{\text{max}}\right)\approx \frac{1.45\delta 
    \kappa^2}{ \sqrt{\delta^2+4\kappa^2}}.
\end{equation}
The above equation suggests that for large $\kappa$ values, the maximum 
power increases linearly with increasing $\kappa$. The expression of the 
stored power at charging time is obtained as
\begin{equation}
    P\left(t=T\right)=\frac{4\delta \kappa^2}{\pi 
    \sqrt{\delta^2+4\kappa^2}}\approx \frac{1.27\delta 
    \kappa^2}{\sqrt{\delta^2+4\kappa^2}}.
\end{equation}

\begin{figure}[h]
	\centering
	\includegraphics{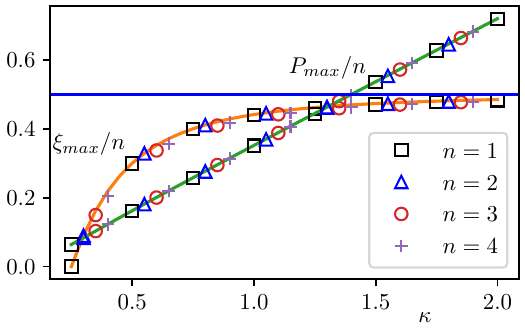}
         \caption{Maximum ergotropy and power, divided by the number of 
         batteries, as a function of the coupling strength $\kappa$. The 
         numerical data points are represented by black squares, blue 
         triangles, red circles, and purple plus signs for one, two, 
         three, and four batteries, respectively. The orange and green 
         solid lines correspond to the analytical formulas for maximum 
         ergotropy and power with one battery, derived in Eq. 
         \ref{eq:max_ergotropy} and \ref{eq:max_power}, respectively. The 
         blue flat line represents optimal work extraction.}
\label{fig:max_erg_vs_kappa}
\end{figure}

Ergotropy attains its maximum value at the charging time $t=T$ and is 
given by,
\begin{equation}\label{eq:max_ergotropy}
    \xi(T)=\delta\left[\frac{4\kappa^2-\delta^2}{4\kappa^2+\delta^2}\right].
\end{equation}
From this equation, it is clear that the maximum ergotropy increases with 
the coupling strength $\kappa$ till $4 \kappa^2 \gtrsim \delta^2$ and 
eventually saturates to the value $\delta$ in the limit $\kappa>>\delta$.
This implies that for the sunburst quantum Ising battery, optimal work 
extraction is achievable by increasing the coupling strength $\kappa$. 
Optimal work extraction here refers to a situation where all stored energy 
in the battery can be extracted without any waste of energy 
\cite{Liu_2021}. With Eqs. \ref{eq:max_power}  and \ref{eq:max_ergotropy}, 
we establish that for the sunburst quantum Ising battery, both the 
power and maximum value of ergotropy increase with increasing the 
coupling strength and optimal work extraction is possible in the strong 
coupling limit (see Fig. \ref{fig:max_erg_vs_kappa}). In contrast, for 
other models like spin chains and Dicke quantum batteries, maximum power 
can be increased in the strong interaction limit at the expense of lower 
stored energy (or ergotropy)\cite{Le_2018,Ferraro_2018}.
The observation that the strong coupling limit is not suitable for 
efficient energy exchange between a battery and a charger has been shown 
by studying the 
energy transfer between two-level systems, a two-level system and quantum 
harmonic oscillator, and two quantum harmonic oscillators 
\cite{Andolina_2018}. For the central spin model, the analytical 
evaluation of the maximum ergotropy for two battery cells shows that it 
is independent of the coupling strength \cite{Liu_2021}. In this case, 
optimization is possible only by varying the initial state of the 
charger. For the sunburst quantum Ising battery, we observe in Fig. 
\ref{fig:max_erg_vs_kappa} that even for more than one battery, both the 
maximum power and ergotropy increase with the coupling strength $\kappa$. 
This indicates that simultaneous optimization of both ergotropy and power 
is possible in the strong coupling limit. In the next section, we study 
the initial state dependence of ergotropy in the sunburst quantum Ising 
battery.

\section{Initial state independence}\label{sec:in_st_independence}

The initial studies on quantum batteries establish that unitary operations 
generating more entanglement between battery cells help in speeding up the 
charging time of a quantum battery \cite{Binder_2015, Campaioli_2017, 
Hovhannisyan_2013}. On the other hand, for the Tavis-Cummings model, the 
entanglement between the battery and charger is shown to have a negative 
effect on ergotropy. This negative effect on ergotropy can be reduced to 
some extent by optimization over the initial state of the charger. In 
particular, it has been shown that the initial states of the charger that 
generate more entanglement between the battery and charger are unsuitable 
for optimizing ergotropy. In contrast, semiclassical initial states that 
induce little entanglement can produce optimal ergotropy 
\cite{Andolina_2019}. The initial state dependence is also found for a 
quantum battery modeled by the $N$ number of two-level systems driven by a 
time-dependent classical source \cite{Crescente_2020}. For the central 
spin battery, it has been shown that optimal work extraction is possible 
only if the initial state of the charger is chosen in such a way that the 
number of spin-up charging units for the charger is $m=\frac{N+n}{2}$, 
where $N$ is the number of chargers.

In contrast to other models, the ergotropy of the sunburst quantum ising 
battery is found to be independent of the charger's initial state. 
This is not difficult to understand if the initial state of the charger 
is 
an eigenstate. The Hamiltonian of the charger ($H_c$) almost commutes 
with 
both the Hamiltonian of the battery ($H_b$) and the interaction 
Hamiltonian $V_{cb}$, considering the limit $h \approx 0$. In this 
limit, 
the time evolution operator can be expressed as:
\begin{equation}\label{eq:time_evolution_operator}
    U=\exp(-iHt)\approx \exp[-it(H_b+V_{cb})]\exp(-iH_c t) ,
\end{equation}
where we assume $\left[H_c,H_b+V_{cb}\right] \approx 0$. If the initial 
state of the charger is an eigenstate, it only produces a global phase 
factor, and then naturally, the ergotropy is the same for any such 
eigenstate.

\begin{figure}[h]
    \centering
    \includegraphics{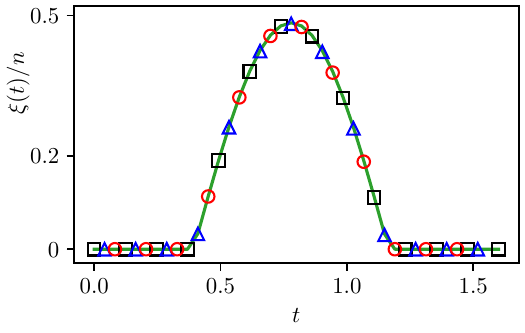}
    \caption{Ergotropy as a function of time for three different 
    realizations of random states (shown by different symbols) as initial 
    state of the charger is plotted here for $L=11, n=1$. The solid line 
    is the analytical curve obtained in Eq. \ref{eq:erg_exp_final}.}
    \label{fig:Initial_state_independence}
\end{figure}

The situation becomes more interesting when the initial state is not 
just an eigenstate. In that case, the operator $\exp(-iH_c t)$ does not 
simply produce a global phase. We numerically calculate the ergotropy by 
considering the initial state of the charger as three different 
realizations of a random state, while the initial state for the battery 
is 
the same as earlier (see Fig. \ref{fig:Initial_state_independence}). We 
observe that the ergotropy turns out to be the same for all three 
different random initial states. This numerically establishes the 
initial 
state independence of ergotropy in the sunburst quantum Ising battery 
case. In this case, charging power is closely related to ergotropy (Eq. 
\ref{eq:power_def}); therefore, not only ergotropy but charging power 
also 
comes out to be independent of the initial state of the charger.

\section{Summary}\label{sec:summary}
In this paper, we study the energy transfer process of the sunburst 
quantum Ising model by treating the transverse field Ising model as an 
extended charger in the limit $J >> h$, with the external qubits 
considered as batteries. We analytically derive expressions for both 
ergotropy and charging power in the case of a single battery, and these 
expressions are found to be in exact agreement with numerical results 
based on exact diagonalization. For the case of two batteries, we 
analytically show that both ergotropy and charging power are exactly twice 
that of a single battery for any value of the interaction strength 
$\kappa$. For more than two batteries, based on numerical results, we 
conjecture that both ergotropy and charging power are linearly 
proportional to the number of batteries. Furthermore, for the sunburst 
quantum Ising battery, we find that the charging power per unit battery is 
$\sqrt{n}$ times larger than the central spin battery \cite{Liu_2021}. 
Like the central spin battery, the charging time is independent of 
the number of batteries, although it can be significantly reduced in the 
strong coupling limit. In contrast to other quantum battery models, we 
observe that both the maximum ergotropy and charging power increase with 
the coupling strength, $\kappa$ \cite{Ferraro_2018, Andolina_2018, 
Le_2018}. We further show that the ergotropy of the sunburst quantum 
Ising battery is independent of the charger's initial state, making it 
more experiment-friendly. Analytical understanding of linear scaling of 
ergotropy and maximum power with the number of batteries remains open.

\appendix
\section{Calculation of stored energy and ergotropy for sunburst quantum Ising model with two batteries }\label{ap:two_qubits}

For sunburst quantum Ising battery for $n=2$, the Hamiltonian of  
batteries is given as $H_b=-\frac{\delta}{2} \left(\sigma_z \otimes 
\unity+\unity \otimes \sigma_z\right)$, which is a diagonal matrix with 
the diagonal elements $\text{diag}\left(-\delta, 0, 0,\delta\right)$ 
with 
respect to be basis vectors for the batteries, given as 
$\{|00\rangle,|01\rangle,|10\rangle,|11\rangle\}$. This implies that we 
can compute the exact expression for both the stored energy and 
ergotropy 
with only the first and last diagonal elements of the reduced density 
matrix for the two batteries. To this end, we first note that our 
initial 
state is chosen as $\kt{\psi(0^-)} = \kt{\psi_G^I} \otimes \kt{00}$. As 
explained in the main text, the time evolution operator in the limit 
$J>>h$ can be expressed as Eq.\ref{eq:time_evolution_operator}.  The 
second term in the time evolution operator will generate a global phase 
during the time evolution, and hence, its effect can be ignored. As a 
result, the time-evolved state can be obtained by repeated applications 
of 
both the Hamiltonian $H_b$ and $V_{cb}$ on the initial state. Since we 
are only interested in computing the first and last diagonal elements of 
the reduced density matrix for the two batteries, we need to calculate 
the 
coefficients of the two states $|\psi_G^I\rangle \otimes |00\rangle$ 
(say 
$A(t)$) and $|\psi_G^I\rangle \otimes |11\rangle$ (say $B(t)$). For ease 
of calculation, we express the terms associated with even powers of time 
separately from those associated with odd powers of time. To start, we 
note down the coefficient of the state $|\psi_G^I\rangle \otimes 
|00\rangle$ associated with odd powers of time, represented as 
$A_{\text{odd}}(t)$.

\begin{equation*}
    \begin{aligned}
        iA_{\textnormal{odd}}(t)= -\delta t 
        +\frac{t^3}{3!}\left[4\delta\kappa^2 
        +\delta^3\right]-\frac{t^5}{5!}\left[16\delta\kappa^4 
        +8\kappa^2\delta^3+\delta^5\right]\\+\frac{t^7}{7!}\left[64\delta\kappa^6
         +48\delta^3\kappa^4+12\kappa^2\delta^5+\delta^7\right]-\cdots
    \end{aligned}
\end{equation*}
With few lines of simplification, the above term can be expressed as
\begin{equation}
    A_{\textnormal{odd}}(t)=i\frac{\delta}{\omega}\sin(\omega t),
\end{equation}
with $\omega=\sqrt{\delta^2+4\kappa^2}$. Now we collect the coefficient which will appear with even order of time:
\begin{equation*}
\begin{aligned}
     A_{\textnormal{even}}(t)=1-\frac{t^2}{2!}\left[2\kappa^2+\delta^2\right]
	+\frac{t^4}{4!}\left[\delta^4+8\kappa^4+6\delta^2\kappa^2\right]\\
	-\frac{t^6}{6!}\left[\delta^6+32\kappa^6+32\delta^2\kappa^4
	+10\delta^4\kappa^2\right]+\cdots .
\end{aligned}
\end{equation*}
The above term can be simplified to
\begin{equation}
    A_{\textnormal{even}}(t)=\cos(\omega t) +\frac{4 \kappa^2}{\omega^2} \sin^2\left(\frac{\omega t}{2}\right).
\end{equation}
The first diagonal element of the reduced density matrix for the two batteries is
\begin{equation}
\begin{aligned}
\lambda_1&=\left|A_{\textnormal{even}}\right|^2+\left|A_{\textnormal{odd}}\right|^2=\cos^2(\omega t)+\frac{16 \kappa^4}{\omega ^4} \sin ^4 \left(\frac{\omega t}{2}\right)
\\&+\frac{8 \kappa^2}{\omega^2} \cos(\omega t) \sin^2\left(\frac{\omega t}{2}\right)+\frac{\delta^2}{\omega^2} \sin^2\left(\omega t\right).
\end{aligned}
\end{equation}
It can be shown that for the state $|\psi_G^I\rangle \otimes |11\rangle$, there is no term associated with odd powers of time $t$, i.e., $B_{\text{odd}}(t) = 0$. $B_{\text{even}}(t)$ can be expressed as
\begin{equation}
B_{\textnormal{even}}(t)=-\frac{4 \kappa^2}{\omega ^2} \sin ^2 \left(\frac{\omega t}{2}\right).
\end{equation}
The fourth diagonal element of the reduced density matrix for the two batteries is
\begin{equation}
\lambda_4=\left|B_{\textnormal{even}}\right|^2+\left|B_{\textnormal{odd}}\right|^2 
=\frac{16 \kappa^4}{\omega ^4} \sin ^4 \left(\frac{\omega t}{2}\right)
\end{equation}
The total amount of stored energy for the two batteries at time $t$ can be expressed as
\begin{equation}\label{eq:stored_eng_two_qubits}
    \Delta E(t)= \delta(1+\lambda_4-\lambda_1)
    = \frac{4\delta \kappa^2}{\omega^2}\left(1-\cos(\omega t)\right).
\end{equation}
From Eq. \ref{eq:stored_eng} and Eq. \ref{eq:stored_eng_two_qubits}, it is 
evident that for two batteries, the stored energy for all values of 
interaction strength $\kappa$  is exactly twice that of a single battery 
at all times $t$. For two batteries, the time at which the stored energy 
becomes maximum exactly coincides with the charging time of a single 
battery (Eq. \ref{eq:charging_time}). The ergotropy becomes non-zero at 
time $t=t^*$, which is obtained by setting $|A(T)|^2$ and $|B(T)|^2$ as 
equal. For two batteries, $t^*$ comes out to be the same as the single 
battery 
(Eq.\ref{eq:charging_begin_time}). The non-zero ergotropy can be obtained 
as
\begin{equation}
    \xi(t)=2\delta\left[\left(\frac{8\kappa^2}{4\kappa^2+\delta^2}\right)\sin^2\frac{\omega t}{2}-1\right],
\end{equation}
which is again exactly twice that of a single battery at all times $t$ 
and for all interaction strength values $\kappa$.

\twocolumngrid
\bibliographystyle{unsrtnat}
\bibliography{references_ergotropy}

\begin{thebibliography}{37}
\providecommand{\natexlab}[1]{#1}
\providecommand{\url}[1]{\texttt{#1}}
\expandafter\ifx\csname urlstyle\endcsname\relax
  \providecommand{\doi}[1]{doi: #1}\else
  \providecommand{\doi}{doi: \begingroup \urlstyle{rm}\Url}\fi

\bibitem[Alicki and Fannes(2013)]{Alicki_2013}
Robert Alicki and Mark Fannes.
\newblock Entanglement boost for extractable work from ensembles of quantum
  batteries.
\newblock \emph{Phys. Rev. E}, 87:\penalty0 042123, Apr 2013.
\newblock \doi{10.1103/PhysRevE.87.042123}.
\newblock URL \url{https://link.aps.org/doi/10.1103/PhysRevE.87.042123}.

\bibitem[Binder et~al.(2015)Binder, Vinjanampathy, Modi, and
  Goold]{Binder_2015}
Felix~C Binder, Sai Vinjanampathy, Kavan Modi, and John Goold.
\newblock Quantacell: powerful charging of quantum batteries.
\newblock \emph{New Journal of Physics}, 17\penalty0 (7):\penalty0 075015, jul
  2015.
\newblock \doi{10.1088/1367-2630/17/7/075015}.
\newblock URL \url{https://dx.doi.org/10.1088/1367-2630/17/7/075015}.

\bibitem[Campaioli et~al.(2017)Campaioli, Pollock, Binder, C{\'e}leri, Goold,
  Vinjanampathy, and Modi]{Campaioli_2017}
Francesco Campaioli, {Felix A} Pollock, {Felix C} Binder, Lucas C{\'e}leri,
  John Goold, Sai Vinjanampathy, and Kavan Modi.
\newblock Enhancing the charging power of quantum batteries.
\newblock \emph{Physical Review Letters}, 118\penalty0 (15), April 2017.
\newblock ISSN 0031-9007.
\newblock \doi{10.1103/PhysRevLett.118.150601}.

\bibitem[Hovhannisyan et~al.(2013)Hovhannisyan, Perarnau-Llobet, Huber, and
  Ac\'{\i}n]{Hovhannisyan_2013}
Karen~V. Hovhannisyan, Mart\'{\i} Perarnau-Llobet, Marcus Huber, and Antonio
  Ac\'{\i}n.
\newblock Entanglement generation is not necessary for optimal work extraction.
\newblock \emph{Phys. Rev. Lett.}, 111:\penalty0 240401, Dec 2013.
\newblock \doi{10.1103/PhysRevLett.111.240401}.
\newblock URL \url{https://link.aps.org/doi/10.1103/PhysRevLett.111.240401}.

\bibitem[Andolina et~al.(2019)Andolina, Keck, Mari, Campisi, Giovannetti, and
  Polini]{Andolina_2019}
Gian~Marcello Andolina, Maximilian Keck, Andrea Mari, Michele Campisi, Vittorio
  Giovannetti, and Marco Polini.
\newblock Extractable work, the role of correlations, and asymptotic freedom in
  quantum batteries.
\newblock \emph{Phys. Rev. Lett.}, 122:\penalty0 047702, Feb 2019.
\newblock \doi{10.1103/PhysRevLett.122.047702}.
\newblock URL \url{https://link.aps.org/doi/10.1103/PhysRevLett.122.047702}.

\bibitem[Pirmoradian and M\o{}lmer(2019)]{Pirmoradin_2019}
Faezeh Pirmoradian and Klaus M\o{}lmer.
\newblock Aging of a quantum battery.
\newblock \emph{Phys. Rev. A}, 100:\penalty0 043833, Oct 2019.
\newblock \doi{10.1103/PhysRevA.100.043833}.
\newblock URL \url{https://link.aps.org/doi/10.1103/PhysRevA.100.043833}.

\bibitem[Ferraro et~al.(2018)Ferraro, Campisi, Andolina, Pellegrini, and
  Polini]{Ferraro_2018}
Dario Ferraro, Michele Campisi, Gian~Marcello Andolina, Vittorio Pellegrini,
  and Marco Polini.
\newblock High-power collective charging of a solid-state quantum battery.
\newblock \emph{Phys. Rev. Lett.}, 120:\penalty0 117702, Mar 2018.
\newblock \doi{10.1103/PhysRevLett.120.117702}.
\newblock URL \url{https://link.aps.org/doi/10.1103/PhysRevLett.120.117702}.

\bibitem[Crescente et~al.(2020)Crescente, Carrega, Sassetti, and
  Ferraro]{Crescente_2020}
Alba Crescente, Matteo Carrega, Maura Sassetti, and Dario Ferraro.
\newblock Ultrafast charging in a two-photon dicke quantum battery.
\newblock \emph{Phys. Rev. B}, 102:\penalty0 245407, Dec 2020.
\newblock \doi{10.1103/PhysRevB.102.245407}.
\newblock URL \url{https://link.aps.org/doi/10.1103/PhysRevB.102.245407}.

\bibitem[Zhang and Blaauboer(2023)]{Zhang_2023}
Xiang Zhang and Miriam Blaauboer.
\newblock Enhanced energy transfer in a dicke quantum battery.
\newblock \emph{Frontiers in Physics}, 10, January 2023.
\newblock ISSN 2296-424X.
\newblock \doi{10.3389/fphy.2022.1097564}.
\newblock URL \url{http://dx.doi.org/10.3389/fphy.2022.1097564}.

\bibitem[Andolina et~al.(2018{\natexlab{a}})Andolina, Farina, Mari, Pellegrini,
  Giovannetti, and Polini]{Marcello_2018}
Gian~Marcello Andolina, Donato Farina, Andrea Mari, Vittorio Pellegrini,
  Vittorio Giovannetti, and Marco Polini.
\newblock Charger-mediated energy transfer in exactly solvable models for
  quantum batteries.
\newblock \emph{Phys. Rev. B}, 98:\penalty0 205423, Nov 2018{\natexlab{a}}.
\newblock \doi{10.1103/PhysRevB.98.205423}.
\newblock URL \url{https://link.aps.org/doi/10.1103/PhysRevB.98.205423}.

\bibitem[Le et~al.(2018)Le, Levinsen, Modi, Parish, and Pollock]{Le_2018}
Thao~P. Le, Jesper Levinsen, Kavan Modi, Meera~M. Parish, and Felix~A. Pollock.
\newblock Spin-chain model of a many-body quantum battery.
\newblock \emph{Phys. Rev. A}, 97:\penalty0 022106, Feb 2018.
\newblock \doi{10.1103/PhysRevA.97.022106}.
\newblock URL \url{https://link.aps.org/doi/10.1103/PhysRevA.97.022106}.

\bibitem[Ghosh et~al.(2020)Ghosh, Chanda, and Sen(De)]{Ghosh_2020}
Srijon Ghosh, Titas Chanda, and Aditi Sen(De).
\newblock Enhancement in the performance of a quantum battery by ordered and
  disordered interactions.
\newblock \emph{Phys. Rev. A}, 101:\penalty0 032115, Mar 2020.
\newblock \doi{10.1103/PhysRevA.101.032115}.
\newblock URL \url{https://link.aps.org/doi/10.1103/PhysRevA.101.032115}.

\bibitem[Zhao et~al.(2021)Zhao, Dou, and Zhao]{Zhao_2021}
Fang Zhao, Fu-Quan Dou, and Qing Zhao.
\newblock Quantum battery of interacting spins with environmental noise.
\newblock \emph{Phys. Rev. A}, 103:\penalty0 033715, Mar 2021.
\newblock \doi{10.1103/PhysRevA.103.033715}.
\newblock URL \url{https://link.aps.org/doi/10.1103/PhysRevA.103.033715}.

\bibitem[Rossini et~al.(2019)Rossini, Andolina, and Polini]{Rossini_2019}
Davide Rossini, Gian~Marcello Andolina, and Marco Polini.
\newblock Many-body localized quantum batteries.
\newblock \emph{Phys. Rev. B}, 100:\penalty0 115142, Sep 2019.
\newblock \doi{10.1103/PhysRevB.100.115142}.
\newblock URL \url{https://link.aps.org/doi/10.1103/PhysRevB.100.115142}.

\bibitem[Ghosh et~al.(2021)Ghosh, Chanda, Mal, and Sen(De)]{Ghosh_2021}
Srijon Ghosh, Titas Chanda, Shiladitya Mal, and Aditi Sen(De).
\newblock Fast charging of a quantum battery assisted by noise.
\newblock \emph{Phys. Rev. A}, 104:\penalty0 032207, Sep 2021.
\newblock \doi{10.1103/PhysRevA.104.032207}.
\newblock URL \url{https://link.aps.org/doi/10.1103/PhysRevA.104.032207}.

\bibitem[Zhang et~al.(2019)Zhang, Yang, Fu, and Wang]{Zhang_2019}
Yu-Yu Zhang, Tian-Ran Yang, Libin Fu, and Xiaoguang Wang.
\newblock Powerful harmonic charging in a quantum battery.
\newblock \emph{Phys. Rev. E}, 99:\penalty0 052106, May 2019.
\newblock \doi{10.1103/PhysRevE.99.052106}.
\newblock URL \url{https://link.aps.org/doi/10.1103/PhysRevE.99.052106}.

\bibitem[Liu et~al.(2021)Liu, Shi, Shi, Wang, and Yang]{Liu_2021}
Jia-Xuan Liu, Hai-Long Shi, Yun-Hao Shi, Xiao-Hui Wang, and Wen-Li Yang.
\newblock Entanglement and work extraction in the central-spin quantum battery.
\newblock \emph{Phys. Rev. B}, 104:\penalty0 245418, Dec 2021.
\newblock \doi{10.1103/PhysRevB.104.245418}.
\newblock URL \url{https://link.aps.org/doi/10.1103/PhysRevB.104.245418}.

\bibitem[Peng et~al.(2021)Peng, He, Chesi, Lin, and Guan]{Peng_2021}
Li~Peng, Wen-Bin He, Stefano Chesi, Hai-Qing Lin, and Xi-Wen Guan.
\newblock Lower and upper bounds of quantum battery power in multiple central
  spin systems.
\newblock \emph{Phys. Rev. A}, 103:\penalty0 052220, May 2021.
\newblock \doi{10.1103/PhysRevA.103.052220}.
\newblock URL \url{https://link.aps.org/doi/10.1103/PhysRevA.103.052220}.

\bibitem[Hu et~al.(2022)Hu, Qiu, Souza, Yuan, Zhou, Zhang, Chu, Pan, Hu, Li,
  Xu, Zhong, Liu, Yan, Tan, Bachelard, Villas-Boas, Santos, and Yu]{Hu_2022}
Chang-Kang Hu, Jiawei Qiu, Paulo J~P Souza, Jiahao Yuan, Yuxuan Zhou, Libo
  Zhang, Ji~Chu, Xianchuang Pan, Ling Hu, Jian Li, Yuan Xu, Youpeng Zhong, Song
  Liu, Fei Yan, Dian Tan, R~Bachelard, C~J Villas-Boas, Alan~C Santos, and
  Dapeng Yu.
\newblock Optimal charging of a superconducting quantum battery.
\newblock \emph{Quantum Science and Technology}, 7\penalty0 (4):\penalty0
  045018, aug 2022.
\newblock \doi{10.1088/2058-9565/ac8444}.
\newblock URL \url{https://doi.org/10.1088}.

\bibitem[Huang et~al.(2023)Huang, Wang, Xiao, Gao, Lin, and Xue]{HuaWan2023}
Xiaojian Huang, Kunkun Wang, Lei Xiao, Lei Gao, Haiqing Lin, and Peng Xue.
\newblock Demonstration of the charging progress of quantum batteries.
\newblock \emph{Phys. Rev. A}, 107:\penalty0 L030201, Mar 2023.
\newblock \doi{10.1103/PhysRevA.107.L030201}.
\newblock URL \url{https://link.aps.org/doi/10.1103/PhysRevA.107.L030201}.

\bibitem[Arjmandi et~al.(2023)Arjmandi, Mohammadi, Saguia, Sarandy, and
  Santos]{ArjMohSantos2023}
Mohammad~B. Arjmandi, Hamidreza Mohammadi, Andreia Saguia, Marcelo~S. Sarandy,
  and Alan~C. Santos.
\newblock Localization effects in disordered quantum batteries.
\newblock \emph{Phys. Rev. E}, 108:\penalty0 064106, Dec 2023.
\newblock \doi{10.1103/PhysRevE.108.064106}.
\newblock URL \url{https://link.aps.org/doi/10.1103/PhysRevE.108.064106}.

\bibitem[Joshi and Mahesh(2022)]{Joshi_2022}
Jitendra Joshi and T.~S. Mahesh.
\newblock Experimental investigation of a quantum battery using star-topology
  nmr spin systems.
\newblock \emph{Phys. Rev. A}, 106:\penalty0 042601, Oct 2022.
\newblock \doi{10.1103/PhysRevA.106.042601}.
\newblock URL \url{https://link.aps.org/doi/10.1103/PhysRevA.106.042601}.

\bibitem[{Allahverdyan, A. E.} et~al.(2004){Allahverdyan, A. E.}, {Balian, R.},
  and {Nieuwenhuizen, Th. M.}]{Balian_2004}
{Allahverdyan, A. E.}, {Balian, R.}, and {Nieuwenhuizen, Th. M.}
\newblock Maximal work extraction from finite quantum systems.
\newblock \emph{Europhys. Lett.}, 67\penalty0 (4):\penalty0 565--571, 2004.
\newblock \doi{10.1209/epl/i2004-10101-2}.
\newblock URL \url{https://doi.org/10.1209/epl/i2004-10101-2}.

\bibitem[Andolina et~al.(2018{\natexlab{b}})Andolina, Farina, Mari, Pellegrini,
  Giovannetti, and Polini]{Andolina_2018}
Gian~Marcello Andolina, Donato Farina, Andrea Mari, Vittorio Pellegrini,
  Vittorio Giovannetti, and Marco Polini.
\newblock Charger-mediated energy transfer in exactly solvable models for
  quantum batteries.
\newblock \emph{Phys. Rev. B}, 98:\penalty0 205423, Nov 2018{\natexlab{b}}.
\newblock \doi{10.1103/PhysRevB.98.205423}.
\newblock URL \url{https://link.aps.org/doi/10.1103/PhysRevB.98.205423}.

\bibitem[Franchi et~al.(2022{\natexlab{a}})Franchi, Rossini, and
  Vicari]{Franchi_2022}
Alessio Franchi, Davide Rossini, and Ettore Vicari.
\newblock Quantum many-body spin rings coupled to ancillary spins: The sunburst
  quantum ising model.
\newblock \emph{Phys. Rev. E}, 105:\penalty0 054111, May 2022{\natexlab{a}}.
\newblock \doi{10.1103/PhysRevE.105.054111}.
\newblock URL \url{https://link.aps.org/doi/10.1103/PhysRevE.105.054111}.

\bibitem[Franchi et~al.(2022{\natexlab{b}})Franchi, Rossini, and
  Vicari]{Rossini_2022}
Alessio Franchi, Davide Rossini, and Ettore Vicari.
\newblock Decoherence and energy flow in the sunburst quantum ising model.
\newblock \emph{Journal of Statistical Mechanics: Theory and Experiment},
  2022\penalty0 (8):\penalty0 083103, aug 2022{\natexlab{b}}.
\newblock \doi{10.1088/1742-5468/ac8284}.
\newblock URL \url{https://dx.doi.org/10.1088/1742-5468/ac8284}.

\bibitem[Mitra and Srivastava(2023)]{Akashash_2023}
Akash Mitra and Shashi C.~L. Srivastava.
\newblock Sunburst quantum ising model under interaction quench: Entanglement
  and role of initial state coherence.
\newblock \emph{Phys. Rev. E}, 108:\penalty0 054114, Nov 2023.
\newblock \doi{10.1103/PhysRevE.108.054114}.
\newblock URL \url{https://link.aps.org/doi/10.1103/PhysRevE.108.054114}.

\bibitem[Villazon et~al.(2020)Villazon, Chandran, and Claeys]{Villazon_2020}
Tamiro Villazon, Anushya Chandran, and Pieter~W. Claeys.
\newblock Integrability and dark states in an anisotropic central spin model.
\newblock \emph{Phys. Rev. Res.}, 2:\penalty0 032052, Aug 2020.
\newblock \doi{10.1103/PhysRevResearch.2.032052}.
\newblock URL \url{https://link.aps.org/doi/10.1103/PhysRevResearch.2.032052}.

\bibitem[Schr\"odinger(1935)]{Schrodinger1935}
E.~Schr\"odinger.
\newblock Die gegenw\"artige situation in der quantenmechanik.
\newblock \emph{Naturwissenschaften}, 23\penalty0 (48):\penalty0 807--812,
  1935.
\newblock ISSN 1432-1904.
\newblock \doi{10.1007/BF01491891}.
\newblock URL \url{https://doi.org/10.1007/BF01491891}.

\bibitem[Greenberger et~al.(1989)Greenberger, Horne, and
  Zeilinger]{Greenberger1989}
Daniel~M. Greenberger, Michael~A. Horne, and Anton Zeilinger.
\newblock \emph{Going Beyond Bell's Theorem}, pages 69--72.
\newblock Springer Netherlands, Dordrecht, 1989.

\bibitem[Islam et~al.(2015)Islam, Ma, Preiss, Eric~Tai, Lukin, Rispoli, and
  Greiner]{islam2015}
Rajibul Islam, Ruichao Ma, Philipp~M Preiss, M~Eric~Tai, Alexander Lukin,
  Matthew Rispoli, and Markus Greiner.
\newblock Measuring entanglement entropy in a quantum many-body system.
\newblock \emph{Nature}, 528\penalty0 (7580):\penalty0 77--83, 2015.

\bibitem[Kaufman et~al.(2016)Kaufman, Tai, Lukin, Rispoli, Schittko, Preiss,
  and Greiner]{kaufman2016}
Adam~M Kaufman, M~Eric Tai, Alexander Lukin, Matthew Rispoli, Robert Schittko,
  Philipp~M Preiss, and Markus Greiner.
\newblock Quantum thermalization through entanglement in an isolated many-body
  system.
\newblock \emph{Science}, 353\penalty0 (6301):\penalty0 794--800, 2016.

\bibitem[Linke et~al.(2018)Linke, Johri, Figgatt, Landsman, Matsuura, and
  Monroe]{linke2018measuring}
Norbert~M Linke, Sonika Johri, Caroline Figgatt, Kevin~A Landsman, Anne~Y
  Matsuura, and Christopher Monroe.
\newblock Measuring the r{\'e}nyi entropy of a two-site fermi-hubbard model on
  a trapped ion quantum computer.
\newblock \emph{Physical Review A}, 98\penalty0 (5):\penalty0 052334, 2018.

\bibitem[Brydges et~al.(2019)Brydges, Elben, Jurcevic, Vermersch, Maier,
  Lanyon, Zoller, Blatt, and Roos]{Roos2019renyimeasure}
Tiff Brydges, Andreas Elben, Petar Jurcevic, Benoît Vermersch, Christine
  Maier, Ben~P. Lanyon, Peter Zoller, Rainer Blatt, and Christian~F. Roos.
\newblock Probing rényi entanglement entropy via randomized measurements.
\newblock \emph{Science}, 364\penalty0 (6437):\penalty0 260--263, 2019.
\newblock \doi{10.1126/science.aau4963}.
\newblock URL \url{https://www.science.org/doi/abs/10.1126/science.aau4963}.

\bibitem[Pusz and Woronowicz(1978)]{Pusz1978}
W.~Pusz and S.~L. Woronowicz.
\newblock Passive states and kms states for general quantum systems.
\newblock \emph{Communications in Mathematical Physics}, 58\penalty0
  (3):\penalty0 273--290, 1978.
\newblock ISSN 1432-0916.
\newblock \doi{10.1007/BF01614224}.
\newblock URL \url{https://doi.org/10.1007/BF01614224}.

\bibitem[Lenard(1978)]{Lenard1978}
A.~Lenard.
\newblock Thermodynamical proof of the gibbs formula for elementary quantum
  systems.
\newblock \emph{Journal of Statistical Physics}, 19\penalty0 (6):\penalty0
  575--586, 1978.
\newblock ISSN 1572-9613.
\newblock \doi{10.1007/BF01011769}.
\newblock URL \url{https://doi.org/10.1007/BF01011769}.

\bibitem[Kamin et~al.(2021)Kamin, Salimi, and Santos]{Kamin_2021}
F.~H. Kamin, S.~Salimi, and Alan~C. Santos.
\newblock Exergy of passive states: Waste energy after ergotropy extraction.
\newblock \emph{Phys. Rev. E}, 104:\penalty0 034134, Sep 2021.
\newblock \doi{10.1103/PhysRevE.104.034134}.
\newblock URL \url{https://link.aps.org/doi/10.1103/PhysRevE.104.034134}.

\end{thebibliography}
\end{document}